\documentclass[12pt]{article}
\usepackage{amsmath,amssymb,amsthm}
\usepackage{mathbbol,bbm}
\usepackage[final]{showkeys}
\usepackage{bm}
\usepackage{calc}

\newcommand{\Cx}{{\mathbb C}}

\newcommand{\idty}{\mathbb{1}}

 \DeclareMathOperator{\tr}{Tr}
\newcommand{\<}{\langle}
\renewcommand{\>}{\rangle}
\providecommand{\abs}[1]{\lvert#1\rvert}
\providecommand{\norm}[1]{\lVert#1\rVert}

\renewcommand{\c}[1]{\mathcal{#1}}

\renewcommand{\r}[1]{\mathrm{#1}}

\begin{document}

\begin{center}
{\LARGE Decay of fidelity in terms of correlation functions} \\[18pt]
R.~Alicki$^\dagger$ and M.~Fannes$^\ddagger$ \\[6pt]
$^\dagger$ Institute of Theoretical Physics and Astrophysics \\
University of Gda\'nsk, Poland \\[6pt]
$^\ddagger$ Instituut voor Theoretische Fysica \\
K.U.Leuven, Belgium
\end{center}

\begin{abstract}
We consider, within the algebraic formalism, the time dependence of fidelity for qubits encoded into an open physical system. We relate the decay of fidelity to the evolution of correlation functions and, in the particular case of a Markovian dynamics, to the spectral gap of the generator of the semigroup. The results are applicable to the analysis of models of quantum memories.
\end{abstract}

\section{Introduction}

Since the advent of quantum information theory it was clear that the particular quantum features relevant for quantum information processing like superpositions of easily distinguishable states and entanglement of well-separated subsystems are extremely fragile with respect to an interaction with the environment \cite{Shor:95, Knill:98}. More recent developments show that the only possible solution to this problem is to encode quantum qubits as fictitious subsystems of real physical systems described by observables that are robust or, in other words, metastable with respect to an external noise \cite{Lidar}. Mathematically, this implies that the algebra $\c A$ of observables of the total system which consists of the relevant part of the Universe has a tensor product structure $\c A = \c Q \otimes \c B$, where $\c Q$ is a finite dimensional matrix algebra describing a single or several encoded qubits. The time evolution is defined on the level of the total system and in the following we consider two cases: 
\begin{enumerate}
\item[a)]
The algebra $\c A$ is a model of the largest relevant isolated system, its dynamics is reversible and described in the Heisenberg picture by a family of automorphisms, i.e.\ unitary maps. 
\item[b)]
The algebra $\c A$ corresponds to a well-defined and well-controlled, spatially confined quantum system, often consisting of  $N\gg1$ physical qubits, which interacts weakly with the environment. Then the environment can be eliminated and one obtains an irreversible reduced dynamics on $\c A$ given by a family of completely positive (CP) unity preserving maps, typically in Markovian approximation.
\end{enumerate}

In quantum information theory the natural description involves states of the fictitious encoded qubit system alone given by a time dependent reduced density matrix $\rho(t)\in \c Q$. It is assumed that at time $t_0=0$ we are able to prepare the encoded qubits in an arbitrary pure state  $\rho(0) = |\psi\rangle \langle\psi|$. Its deterioration due to external influences is characterized by the time-dependent fidelity
\begin{equation}
F(\psi; t) =  \< \psi \,,\, \rho(t)\psi \>,\enskip  t\ge 0.
\label{0}
\end{equation}
The problem with equation~(\ref{0}) is that there is in general no physically motivated description of $\rho(t)$ in terms of a map acting on the initial state $\rho(0)$. We shall see that a consistent description in operational terms involves time dependent correlation functions of elements of $\c A$. Such objects are natural in quantum statistical mechanics and the formulation of the problem in a more general algebraic language allows to admit field theoretical or/and infinite volume models of physical systems. Moreover, an algebraic formulation is a natural framework to study irreversible dynamics given in terms of CP maps.

\section{Algebraic formulation}

As we shall be concerned with local perturbations with respect to a given reference state, such as a thermal state, we use the setting of von~Neumann algebras~\cite{RB97}. Let $\c A$ denote the von~Neumann algebra of the (bounded, complex) observables of the total system which is assumed to be a tensor product 
\begin{equation}
\c A = \c Q \otimes \c B.
\label{1}
\end{equation} 
Here $\c Q$ is a the finite dimensional sub-algebra of the encoded qubit observables, assumed to be isomorphic to a full $d\times d$ matrix algebra, and $\c B$ is the syndrome system. The single step effective Heisenberg dynamics of the total system is a normal, completely positive, identity preserving map $\Lambda$ on $\c A$ with a cyclic and faithful, normal, invariant state $\omega$. This state represents a stable reference state typically thermal equilibrium. The general theory~\cite{RB97} assures the existence  of the so-called modular automorphism group (MAG) on $\c A$ denoted by $\tau = \{\tau_t \,:\, -\infty < t<\infty\}$. The defining properties of the MAG are expressed by extending the time parameter $t$ to a complex domain.

The reader who is not familiar with this abstract approach can always restrict to the special case where $\c A$ is the algebra of bounded operators $\c{B(H)}$ on a certain Hilbert space $\c H$ and $\omega$ is a density matrix. The abstract notation $\omega(a)$ for the mean value of the observable $a$ in the state $\omega$ corresponds to the usual formula $\tr (\omega a)$. In this case faithfulness means that $\omega$ is strictly positive, it can therefore always be seen as a canonical Gibbs state corresponding to a certain Hamiltonian $H$ at inverse temperature 1, i.e.\ $\omega = \r e^{-H}/\tr \r e^{-H}$. The corresponding Heisenberg picture dynamics $a \to \tau_t(a) := \r e^{itH}\, a\, \r e^{-it H}$ is a MAG in this standard setting. One can easily check that such a MAG is uniquely defined by the state up to an irrelevant time unit. 

\section{Fidelity and correlations}

The restriction of $\omega$ to $\c Q$, denoted by $\omega_{\c Q}$, is determined by a $d\times d$ density matrix which is strictly positive as $\omega_{\c Q}$ is faithful. It is our aim to describe the evolution of fidelity between an initially pure qubit state and its evolution. Any given density matrix $\sigma$ on $\c Q$ can be obtained as a restriction to $\c Q$ of a local perturbation of the reference state $\omega$. Indeed, any $a \in \c A$ such that $\omega(a^\dagger a) = 1$ defines a perturbed state $\omega'$ on $\c A$ 
\begin{equation}
\omega'(b) := \omega(a^\dagger b\, a). 
\label{17}
\end{equation}
We claim that we can always find an $a$ such that
\begin{equation} 
\omega'\bigr|_{\c Q} = \sigma\enskip \text{or, equivalently, that}\enskip \omega(a^\dagger q\,a) = \sigma(q)\enskip \text{for } q \in\c Q.
\label{23}
\end{equation}
In fact, such an $a$ can even be found in $\c Q$, e.g.\ $a = \sigma^{1/2} \omega_{\c Q}^{-1/2}$. We assume from now on that $a$ is chosen in such a way that $\omega(a^\dagger \cdot a)_{\c Q}$ is the pure qubit state $|\psi\> \<\psi|$. The expression~(\ref{0}) of the fidelity now reads
\begin{equation}
F(\psi; \Lambda) = \omega(a^\dagger \Lambda(P_\psi)\,a) = \omega\bigl( \tau_{-i\beta}(a)\, a^\dagger \Lambda(P_\psi) \bigr).
\label{11}
\end{equation}
Here $P_\psi$ denotes the orthogonal projector on $\psi$ tensorized with the identity on $\c B$ and $\tau$ is the MAG of $\omega$. Equation~(\ref{11}) implicitly assumes that $a$ is an analytic element for the MAG. 

It is useful to introduce the scalar product 
\begin{equation}
\< x \,,\, y \>_\omega := \omega(x^\dagger y)
\label{2}
\end{equation}
on $\c A$ and the Hilbert space $\c H_{\omega}$ obtained by completing $\c A$ with respect to the norm defined by~(\ref{2})
\begin{equation}
\norm x_\omega := \< x \,,\, x \>_\omega^{\frac{1}{2}}.
\label{16}
\end{equation}

Next, we introduce operators
\begin{equation}
x := a\, \tau_{i\beta}(a^\dagger) - \idty \enskip\text{and}\enskip y := P_\psi - \omega(P_\psi).
\label{12}
\end{equation}
It is easy to check that both $x$ and $y$ are centred, i.e.\ orthogonal to $\idty$ with respect to the scalar product~(\ref{2})
\begin{equation}
\omega(x) = \omega(y) = 0.
\label{13}
\end{equation}
The fidelity~(\ref{11}) can now be be expressed in terms of the correlation between the observable $x^\dagger$ and the evolved observable $y$ as 
\begin{equation}
F(\psi; \Lambda) = \omega \bigl( x^\dagger \Lambda(y)\bigr) + \omega(P_\psi). 
\label{14}
\end{equation}
In the case a) with $\omega$ being a thermal equilibrium state the dynamics $\Lambda$ coincides with its MAG and therefore the expression~(\ref{14}) with substitution of~(\ref{12}) is a sum of thermal correlation functions or, for field theoretical models, of thermal Green functions~\cite{AGD}.

The form~(\ref{17}) we used to generate an arbitrary state $\sigma$ on $\c Q$, in particular a pure state $|\psi\> \<\psi|$, does  not correspond to a realistic physical preparation procedure. A more precise description would involve the application of quantum operations. Namely, the preparation of the initial state in $\c Q$ is done through interactions with certain quantum devices. Eliminating the degrees of freedom of these preparation devices we obtain a quantum operation which is, in Heisenberg picture, an identity preserving CP map on $\c A$ of the form
\begin{equation}
\Phi(b) = \sum_j a_j^\dagger b\, a_j \enskip\text{with}\enskip \sum_j a_j^\dagger a_j = \idty .
\label{20a}
\end{equation}
Then, the initial state $\omega$ on $\c A$ is transformed to a perturbed state $\omega'$ given by 
\begin{equation}
\omega'(b) = \sum_j \omega(a_j^\dagger b a_j) \enskip\text{with}\enskip 
\sum_j \omega(a_j^\dagger \cdot a_j)\bigr|_{\c Q} = \sigma . 
\label{20}
\end{equation}
Again, for any initial $\sigma$ on $\c Q$ we can find an operation $\Phi$ with Kraus operators $a_j\in \c Q$, e.g.\ using a CP map of the form 
\begin{equation}
\Phi(q) = \tr(\sigma q)\idty.
\label{13a}
\end{equation}
A similar approach to that used in the case~(\ref{17}) can then be followed. The centred observable $x$ is now given by
\begin{equation}
x := \sum_j a_j\, \tau_{i\beta}(a_j^\dagger) - \idty.
\label{12a}
\end{equation}

\bigskip \noindent
\textbf{Remarks} 
\begin{enumerate}
\item[i)] 
The choice of Kraus operators in~(\ref{20a}) depends of the physical implementation of the preparation procedure. Therefore, in general, the $a_j$ don't need to be elements of $\c Q$ and even if they are the observable $x$ is not in $\c Q$ except for the case where $\omega$ is a product state with respect to the tensor structure~(\ref{1}).
\item[ii)] 
The simple choice~(\ref{17}) of the initial perturbed state can be seen in terms of a more realistic preparation procedure~(\ref{20a}) by considering a weight $p<1$. As $p a^{\dagger} a \le \idty$ there exists an operation $\Phi$ with $a_1 = \sqrt pa$ which produces an ensemble of initial perturbed states $\{ \omega_j (\cdot) = \omega (a_j^{\dagger}\cdot a_j)/\omega (a_j^{\dagger}a_j)\}$. Filtering out the state $\omega_1$ from this ensemble provides a physical preparation of the state~(\ref{17}). This should be compared with distillation procedures, see e.g.~\cite{H}. 
\item[iii)] 
The operational prescription of the initial state preparation~(\ref{20a}--\ref{20}) does in general not lead to a reduced dynamics or subdynamics. Such a reduced dynamics can be always formulated (in  Schr\"odinger picture) in terms of an assignment map $\sigma \to \Psi^*(\sigma) = \omega'$ which assigns to a given initial state $\sigma$ of the qubits the initial state $\omega'$ of the total system~\cite{PA}. The standard reduction procedure can then be applied to yield a dynamical map $\Gamma^*(\sigma) = \tr_{\c B}\bigl( \Lambda^*\Psi^*(\sigma) \bigr)$ with the standard partial trace over the environment $\c B$. Here $\Lambda^*$ is the Schr\"odinger picture version of the CP dynamics for the total system. Note that the particular choice~(\ref{13a}) of the preparing operation produces a subdynamics with a product state assignment map $\sigma \to \sigma \otimes \omega\bigr|_{\c B}$.
\end{enumerate}

\section{Estimating decays}

The case~b) single step dynamics mentioned in the Introduction is given by a CP unity preserving map with invariant state $\omega$. In order to estimate the decay of fidelity we apply Schwarz's inequality to the first term on the rhs of~(\ref{14})
\begin{equation}
\bigl| \omega \bigl( x^\dagger \Lambda(y) \bigr) \bigr|  = \bigl| \< x \,,\, \Lambda(y)\>_\beta \bigr| \le  \norm x_\beta\, \norm{\Lambda(y)}_\beta. 
\label{15}
\end{equation}
Therefore, the relevant part is $\norm{\Lambda(y)}_\beta$.

Using 2-positivity of $\Lambda$ and invariance of $\omega$ we obtain
\begin{equation}
\begin{split}
\Bigl| \< x \,,\, \Lambda(y) \>_\omega \Bigr|^2 
&\le \norm x^2_\omega\, \norm{\Lambda(y)}^2_\omega = \norm x^2_\omega\, \omega \Bigl( \bigl(\Lambda(y)\bigr)^\dagger \Lambda(y) \Bigr) \\
&\le \norm x^2_\omega\, \omega \Bigl( \bigl(\Lambda(y^\dagger y)\bigr) \Bigr) = \norm x^2_\omega\, \omega \bigl( y^\dagger y \bigr) \\ 
&= \norm x^2_\omega\, \norm y^2_\omega.  
\end{split}
\label{3}
\end{equation} 
This means that 
\begin{equation}
x \in \c A \mapsto \Lambda(x)
\label{4}
\end{equation}
is a well-defined contraction on $\c H_{\omega}$. Decomposing $\c H_{\omega}$ into a direct sum of $\Cx \idty$ and its orthogonal complement, $\Lambda$ is of the following form
\begin{equation}
\Lambda = \begin{bmatrix} 1 &\< \varphi \,,\, \cdot \>_\omega \\0 &\tilde\Lambda \end{bmatrix}.
\label{5}
\end{equation}
Here $\tilde\Lambda$ is a map on $\idty^\perp$ and $\varphi$ is a vector in $\idty^\perp$. Next, we express that $\Lambda$ is contractive
\begin{equation}
\biggl\Vert \Lambda\, \begin{bmatrix} \alpha \\\eta \end{bmatrix} \biggr\Vert^2_\omega 
= \biggl\Vert \begin{bmatrix} \alpha + \< \varphi \,,\, \eta \>_\omega \\\tilde\Lambda\,\eta \end{bmatrix}\biggr\Vert^2_\omega = \abs{\alpha + \< \varphi \,,\, \eta \>_\omega}^2 + \norm{ \tilde\Lambda\, \eta}^2_\omega. 
\label{9}
\end{equation}
This should be not larger than
\begin{equation}
\biggl\Vert \begin{bmatrix} \alpha \\\eta \end{bmatrix} \biggr \Vert^2_\omega = \abs\alpha^2 + \norm{\eta}^2_\omega 
\label{10}
\end{equation}
for any choice of $\alpha \in \Cx$ and $\eta \in \idty^\perp$. It follows that $\varphi = 0$ and that $\tilde\Lambda$ is contractive. The properties of $\tilde\Lambda$ determine the fidelity decay to its lowest value $\omega(P_\psi)$. This will be illustrated by the example of a Markovian dynamics.

\section{Irreversible Markovian dynamics}

In fact, in a semi-group description of reduced dynamics, one assumes that the dynamics is described by a weakly continuous semi-group $\{\Lambda_t \,:\, t\ge0\}$ of CP identity preserving maps. In this case the map in~(\ref{5}) becomes time dependent. Applying the argument of above for each $t$ separately we find that 
\begin{equation}
\Lambda_t = \begin{bmatrix} 1 &0 \\0 &\tilde\Lambda_t \end{bmatrix}
\label{18}
\end{equation}
where $\{\tilde\Lambda_t \,:\, t \ge 0 \}$ is a weakly continuous semi-group of contractions on $\idty^\perp$. We now impose that $\tilde\Lambda$ is strictly contractive, i.e.\ that there exists a constant $\gamma>0$ such that
\begin{equation}
\norm{\tilde\Lambda_t(x)}_\omega = \norm{\Lambda_t(x)}_\omega \le \r e^{-\gamma t}\, \norm x_\omega,\enskip t \ge 0,\ x \in \idty^\perp.
\label{7}
\end{equation}
Combining~(\ref{14}), (\ref{15}), and (\ref{7}) we obtain our final estimate
\begin{equation}
F(\psi; t) := F(\psi; \Lambda_t) \le \r e^{-\gamma t}\, \norm x_\beta\, \norm y_\beta + \omega(P_\psi)
\label{19}
\end{equation}
with $x$ and $y$ as in~(\ref{12}) or (\ref{12a}).

\bigskip \noindent
\textbf{An example} 

Assume that $\tau_{i\beta}(q) = q$ for all $q \in \c Q$. This particularly simple case applies e.g.\ to the models of quantum memories in~\cite{K} and~\cite{AFH}. In this case, the restriction of $\omega$ to $\c Q$ is just the tracial state. Assume, moreover, that the semi-group satisfies the detailed balance condition~\cite{Alicki:07} in the form
\begin{equation}
\Lambda_t = \tau_t \circ \r e^{tL_{\r dis}}
\label{21}
\end{equation}
where the dissipative part $L_{\r dis}$ of the generator is self-adjoint and commutes with the system dynamics $\tau$. This is e.g.\ the case for generators obtained by Davies's weak coupling procedure, see~\cite{D}. Let $\lambda$ be the spectral gap of $L_{\r{dis}}$ which is equal to the lowest eigenvalue of $-L_{\r{dis}}$ restricted to $\idty^\perp$. Then, for both simplest choices of initial state preparation given by \\ 
(\ref{17}) with $a = P_\psi\, \sqrt d$, or \\
(\ref{13a}) with $\sigma = |\psi\rangle\langle\psi|$ \\
the general estimate~(\ref{19}) becomes
\begin{equation}
F(\psi; t) \le \frac{1}{d} + \r e^{-\lambda t}\, \Bigl( 1 - \frac{1}{d} \Bigr).
\label{22}
\end{equation} 

\section{Conclusions}

We provided a bridge between the notion of fidelity which characterizes the quality of quantum information stored in a noisy environment and the notion of temporal, thermal correlation functions frequently used in statistical mechanics to characterize ergodic properties of large quantum systems. This technique should be helpful in searching for good candidates for quantum memories among the different models of interacting many body systems. This problem is quite important as properly scalable quantum memories are a necessary ingredient for any attempt at large scale quantum information processing.

\bigskip\noindent
\textbf{Acknowledgements}

This work was done while the authors participated to the 2008 MHQP programme at IMS, NUS, Singapore. They are grateful for the stimulating environment and for the warm hospitality extended to them. The valuable input from M.~Horodecki is also acknowledged. This work is partially funded by the Belgian Interuniversity Attraction Poles Programme P6/02 (MF) and supported by the Polish research network LFPPI (RA).

\end{document}